\documentclass[copyright,creativecommons]{eptcs}
\providecommand{\event}{Programming Language Approaches to Concurrency- \\and Communication-cEntric Software (PLACES 2026)}

\usepackage{amsmath}
\usepackage{amssymb}
\usepackage{mathtools}
\usepackage{amsthm}
\usepackage{graphicx}
\usepackage{bbm}
\usepackage{float}
\usepackage{indentfirst}
\usepackage{hyperref}
\usepackage{longtable,tabularx,booktabs,array,ragged2e}
\usepackage{subcaption}
\usepackage{tikz}
\usepackage{ebproof}
\usepackage{placeins}
\usepackage[T1]{fontenc}
\usepackage{algorithm}
\usepackage[noend]{algpseudocode}
\usetikzlibrary{automata,positioning,arrows.meta,calc,matrix}

\title{Generating Local Shields for Decentralised Partially Observable Markov Decision Processes}
\author{Haoran Yang
\institute{University of Oxford}
\email{haoran.yang@balliol.ox.ac.uk}
\and
Nobuko Yoshida
\institute{University of Oxford}
\email{nobuko.yoshida@cs.ox.ac.uk}
}
\def\titlerunning{Generating Local Shields for Decentralised Partially Observable Markov Decision Processes}
\def\authorrunning{Haoran Yang, Nobuko Yoshida}
\begin{document}
\maketitle

\begin{abstract}
Multi-agent systems under partial observation often struggle to maintain safety because each agent's locally chosen action does not, in general, determine the resulting joint action. Shielding addresses this by filtering actions based on the current state, but most existing techniques either assume access to a shared centralised global state or employ memoryless local filters that cannot consider interaction history.

We introduce a shield process algebra with guarded choice and recursion for specifying safe global behaviour in communication-free Decentralised Partially Observable Markov Decision Process (Dec-POMDP) settings. From a shield process, we compile a process automaton, then a global Mealy machine as a safe joint-action filter, and finally project it to local Mealy machines whose states are belief-style subsets of the global Mealy machine states consistent with each agent's observations, and which output per-agent safe action sets.

We implement the pipeline in Rust\footnote{\url{https://gitlab.cs.ox.ac.uk/ug23hy/shield-process-compilation-pipeline/}} and integrate PRISM, the Probabilistic Symbolic Model Checker\footnote{\url{https://www.prismmodelchecker.org}}, to compute best- and worst-case safety probabilities independently of the agents' policies. A multi-agent path-finding case study demonstrates how different shield processes substantially reduce collisions compared to the unshielded baseline while exhibiting varying levels of expressiveness and conservatism.

\end{abstract}


\section{Introduction}

Multi-agent systems are prone to encountering collisions, deadlocks, or livelocks because an individual agent's local action does not, in general, uniquely determine the resulting joint action. Consequently, ensuring that each agent selects a concrete action that is guaranteed to be safe is non-trivial. This challenge is exacerbated in communication-free, Decentralised Partially Observable Markov Decision Process (Dec-POMDP) settings~\cite{intro-decpomdp}, where each agent must choose its action solely on the basis of its own local observations.

One widely used approach to enforcing safety is shielding, in which a shield restricts the agents' choices by providing, at each state, a set of safe actions. Existing approaches include temporal-logic based synthesis~\cite{alshiekh2017safereinforcementlearningshielding}, probabilistic shields~\cite{jansen2020safe_rl_probabilistic_shields}, learned shields~\cite{Melcer2025LearnedShields}, and decentralisation techniques~\cite{elsayedaly2021safemultiagentreinforcementlearning, melcer2022shield_decentralization_neurips, melcer2024shielddecentralization} that derive local shields from a centralised safe-action set. While these methods have proved effective in many settings, they typically rely on assumptions such as access to a centralised global state, the availability of memoryless local filtering, or the sufficiency of constraints expressed purely at the action level. These assumptions limit their applicability in regimes where agents possess only local information and inter-agent communication is disallowed.

Although shields can sometimes be generated automatically, for example, via DFA synthesis from LTL specifications~\cite{alshiekh2017safereinforcementlearningshielding}, such constructions tend to be rigid and conservative. Moreover, they may fail to capture the informational complexity induced by partial observations, wherein each agent's local view reveals only a fragment of the underlying global state.

To solve these problems, we propose a succinct process algebra with guarded choice and recursion. This algebra admits an automata-theoretic interpretation and can be compiled into Mealy-machine-style shields~\cite{6771467}. The compilation pipeline comprises three main steps. First, we translate the process algebra into a process automaton. Second, we construct a global shield Mealy machine that takes the Dec-POMDP state as input and outputs a decentralised representation of safe joint actions consistent with the process automaton. Third, we derive local shield Mealy machines by replacing the Dec-POMDP state input with each agent's partial observation: the local shield Mealy machines' states are belief-style subsets of global Mealy machine states consistent with the agent's observation, thereby enumerating the possible individual safe action sets that remain feasible under decentralised local information. We implement this pipeline in Rust and analyse the resulting models using an integrated PRISM-based workflow.

In this paper, we present the complete pipeline, from motivation to implementation details, and illustrate it through a simplified multi-agent path-finding (MAPF)~\cite{DBLP:journals/corr/abs-1906-08291} case study. MAPF is used here as a case study because it is easy to visualise and has simple collision rules; however, the pipeline applies more generally to communication-free Dec-POMDPs whenever the relevant safety condition can be represented using the state together with a finite counter.

\paragraph{Contributions} (1) A shield process algebra with recursion and guarded choice for decentralised, partially observable, communication-free settings. (2) A compilation pipeline that, given a shield process specification, generates the corresponding shield process automaton and, using transition and observation descriptions, synthesises a global Mealy shield and projects it to local Mealy shields executable from local observations, which in turn export safe action sets for each agent under partial observability. (3) A Rust implementation of this pipeline with PRISM integration, enabling the computation of best- and worst-case safety probabilities without fixing a policy.

\section{Pipeline Explanation}
This section is structured as follows. First, we introduce the preliminaries that underpin our pipeline. Second, we specify the configuration of a simplified multi-agent path-finding (MAPF) problem, which serves as a running example to illustrate the pipeline. Third, we present the process algebra used to describe a shield in terms of global states, thereby providing a global perspective on system behaviour. Fourth, we describe the pipeline itself using a special case of the MAPF problem with ``blind'' agents. In this final part, we construct a process automaton directly from the process algebra, derive a global shield Mealy machine from the process automaton together with the descriptions of initial states and the joint actions that induce state transitions, and subsequently synthesise a local shield Mealy machine for each agent by applying partial observation and belief-state construction to relate local observations to global states.

\subsection{Preliminaries}

A \textit{decentralised partially observable Markov decision process} (\textit{Dec-POMDP}) is a tuple \cite{intro-decpomdp} $\mathcal{M} = (\mathcal{I}, S, (A_i)_{i\in\mathcal{I}}, \delta, (\Omega_i)_{i\in\mathcal{I}}, O, r, \gamma, \rho_0)$, where $\mathcal{I} = \{1,\dots,n\}$ is the set of agents, $S$ is the global state space, and $A_i$ is the action space of agent $i$ (with joint action space $A := \prod_{i\in\mathcal{I}} A_i$). The transition dynamics are given by $\delta(s'\mid s,a)$, the probability of transitioning from state $s$ to state $s'$ under joint action $a$. For each agent $i$, $\Omega_i$ is the observation space (with joint observation space $\Omega := \prod_{i\in\mathcal{I}} \Omega_i$), and $O(o_1,\dots,o_n\mid s',a)$ denotes the probability of joint observation $(o_1,\dots,o_n)$ after executing $a$ and reaching $s'$. The shared reward function is $r(s,a)$, $\gamma \in (0,1]$ is the discount factor, and $\rho_0$ is the initial-state distribution.

A \textit{global history} up to time $t$ is $h^t = (s^0, a^0, \dots, a^{t-1}, s^t)$, and a \textit{local history} for agent $i$ up to time $t$ is $h_i^t = (o_i^0, a_i^0, \dots, a_i^{t-1}, o_i^t)$. A \textit{decentralised stochastic policy} is a tuple $\pi := (\pi_1,\dots,\pi_n)$, where each $\pi_i(a_i\mid I_i^t)$ is a distribution over $A_i$ given the local information $I_i^t$, typically the local history $h_i^t$ or, in the memoryless case, the current observation $o_i^t$ \cite{melcer2024shielddecentralization}. Bounded histories can be encoded into states or observations by forming the product of states or observations with actions. Consequently, in the remainder of the paper, we focus on states and observations without loss of generality.

A \textit{shield} for an agent is a function $\mathcal{D} : S \rightarrow 2^{A}$ that returns the set of safe actions at each state \cite{alshiekh2017safereinforcementlearningshielding,Melcer2025LearnedShields,melcer2024shielddecentralization}. Shields can be represented, for example, as deterministic finite automata (DFAs) or as Mealy machines \cite{elsayedaly2021safemultiagentreinforcementlearning}. In this work, we consider \textit{pre-posed shields} that filter the available actions before the agents select them.

A \textit{Mealy machine} is a tuple $(Q,q_0,\Sigma_I,\Sigma_O,\delta,\lambda)$, where $Q$ is a finite set of states, $q_0\in Q$ is the initial state, $\Sigma_I$ is the input alphabet, $\Sigma_O$ is the output alphabet, $\delta:Q\times\Sigma_I\to Q$ is the transition function, and $\lambda:Q\times\Sigma_I\to\Sigma_O$ is the output function. In this paper, Mealy machines represent shields whose inputs are either global states or local observations, and whose outputs are safe action sets.

\subsection{A (Simplified) Multi-Agent Path-Finding Problem}

We consider a simplified multi-agent path-finding (MAPF) problem \cite{DBLP:journals/corr/abs-1906-08291}. The environment is a fixed $H \times W$ grid populated by $n$ agents. Each grid cell is either an obstacle or a free cell. A free cell may be empty, occupied by an agent, or designated as an agent's target.

Each agent can perform one of five actions: $\{\leftarrow, \rightarrow, \uparrow, \downarrow, \cdot\}$, which correspond to moving left, right, up, down, or remaining stationary. Agents operate under partial observability; each agent's observation $\Omega_i$ includes its local neighbourhood within radius $R$ and a coarse direction-to-target signal in $\{-1,0,1\} \times \{-1,0,1\}$.

For simplicity, we consider only \emph{vertex conflicts}, meaning that no two agents may occupy the same cell after a transition. Figure~\ref{fig:mapf-grid}(a) shows a representative grid configuration with two agents, their targets, obstacles, and an example observation radius and direction-to-target-signal. Figure~\ref{fig:mapf-grid}(b) depicts the action space for each agent, and Figure~\ref{fig:mapf-grid}(c) illustrates a vertex conflict.

\newsavebox{\mapfleftbox}
\newlength{\mapfcolheight}
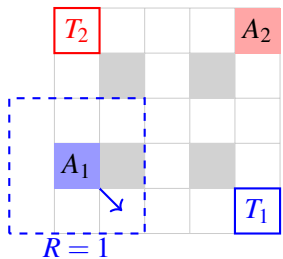
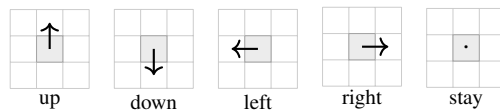
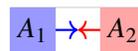
\begin{figure}[H]
  \centering
  \sbox{\mapfleftbox}{%
    \begin{minipage}[t]{0.48\linewidth}
      \vspace{0pt}
      \centering
      \begin{subfigure}[t]{\linewidth}
        \centering
        \begin{tikzpicture}[scale=0.6]
          \draw[step=1cm, gray!40, very thin] (0,0) grid (5,5);

          \fill[gray!35] (1,1) rectangle (2,2);
          \fill[gray!35] (1,3) rectangle (2,4);
          \fill[gray!35] (3,1) rectangle (4,2);
          \fill[gray!35] (3,3) rectangle (4,4);

          \draw[blue, ->, thick] (0.5,1.5) -- (1.5,0.5);

          \fill[blue!40] (0,1) rectangle (1,2);
          \node at (0.5,1.5) {$A_1$};
          \fill[red!35] (4,4) rectangle (5,5);
          \node at (4.5,4.5) {$A_2$};

          \draw[blue, thick] (4,0) rectangle (5,1);
          \node[blue] at (4.5,0.5) {$T_1$};
          \draw[red, thick] (0,4) rectangle (1,5);
          \node[red] at (0.5,4.5) {$T_2$};

          \draw[blue, dashed, thick] (-1,0) rectangle (2,3);
          \node[blue] at (0.5,-0.3) {$R=1$};
        \end{tikzpicture}
        \caption{Grid with agents (filled cells), obstacles (grey cells), targets (bordered cells), and the current observation region of \(A_1\) for \(R=1\).}
      \end{subfigure}
    \end{minipage}
  }
  \setlength{\mapfcolheight}{\ht\mapfleftbox}
  \addtolength{\mapfcolheight}{\dp\mapfleftbox}

  \usebox{\mapfleftbox}
  \hfill
  \begin{minipage}[t][\mapfcolheight]{0.48\linewidth}
    \vspace{0pt}
    \centering
    \begin{subfigure}[t]{\linewidth}
      \centering
      \begin{minipage}{\linewidth}
        \centering
        \begin{tikzpicture}[scale=0.35]
          \draw[step=1cm, gray!40, very thin] (0,0) grid (3,3);
          \fill[gray!15] (1,1) rectangle (2,2);
          \draw[gray!70] (1,1) rectangle (2,2);
          \draw[->, thick] (1.5,1.5) -- (1.5,2.5);
          \node[font=\scriptsize] at (1.5,-0.5) {up};
        \end{tikzpicture}
        \hspace{0.35em}
        \begin{tikzpicture}[scale=0.35]
          \draw[step=1cm, gray!40, very thin] (0,0) grid (3,3);
          \fill[gray!15] (1,1) rectangle (2,2);
          \draw[gray!70] (1,1) rectangle (2,2);
          \draw[->, thick] (1.5,1.5) -- (1.5,0.5);
          \node[font=\scriptsize] at (1.5,-0.5) {down};
        \end{tikzpicture}
        \hspace{0.35em}
        \begin{tikzpicture}[scale=0.35]
          \draw[step=1cm, gray!40, very thin] (0,0) grid (3,3);
          \fill[gray!15] (1,1) rectangle (2,2);
          \draw[gray!70] (1,1) rectangle (2,2);
          \draw[->, thick] (1.5,1.5) -- (0.5,1.5);
          \node[font=\scriptsize] at (1.5,-0.5) {left};
        \end{tikzpicture}
        \hspace{0.35em}
        \begin{tikzpicture}[scale=0.35]
          \draw[step=1cm, gray!40, very thin] (0,0) grid (3,3);
          \fill[gray!15] (1,1) rectangle (2,2);
          \draw[gray!70] (1,1) rectangle (2,2);
          \draw[->, thick] (1.5,1.5) -- (2.5,1.5);
          \node[font=\scriptsize] at (1.5,-0.5) {right};
        \end{tikzpicture}
        \hspace{0.35em}
        \begin{tikzpicture}[scale=0.35]
          \draw[step=1cm, gray!40, very thin] (0,0) grid (3,3);
          \fill[gray!15] (1,1) rectangle (2,2);
          \draw[gray!70] (1,1) rectangle (2,2);
          \node at (1.5,1.5) {$\cdot$};
          \node[font=\scriptsize] at (1.5,-0.5) {stay};
        \end{tikzpicture}
      \end{minipage}
      \caption{Agent action space.}
    \end{subfigure}
    \vspace{0.6em}
    \vfill
    \begin{subfigure}[t]{\linewidth}
      \centering
      \begin{tikzpicture}[scale=0.6]
        \draw[step=1cm, gray!40, very thin] (0,0) grid (3,1);

        \draw[blue, ->, thick] (0.5,0.5) -- (1.5,0.5);
        \draw[red, ->, thick] (2.5,0.5) -- (1.5,0.5);

        \fill[blue!40] (0,0) rectangle (1,1);
        \node at (0.5,0.5) {$A_1$};
        \fill[red!35] (2,0) rectangle (3,1);
        \node at (2.5,0.5) {$A_2$};
      \end{tikzpicture}
      \caption{Vertex conflict example.}
    \end{subfigure}
  \end{minipage}
  \caption{Example MAPF grids.}
  \label{fig:mapf-grid}
\end{figure}

\subsection{Shield Process Algebra}

\paragraph{Shield Process} We define a \emph{global state shield} as a set of states $Sh\subseteq S$ that are considered safe to move next. We also define a \emph{shield process} using the following grammar:

\[
P ::= \textsf{idle}\mid \textsf{fail}\mid\mu X.P\mid X\mid Sh.P\mid P_1\Vert_{g}P_2,
\]

where $\textsf{idle}$ denotes successful termination of the shield process, and $\textsf{fail}$ denotes unsuccessful termination. The construction $\mu X.\,P$ introduces recursion (which is assumed to be guarded), with $X$ as the corresponding recursion variable. The term $Sh.P$ denotes a continuation process that asserts that the next state lies in the global state shield $Sh \subseteq S$ and then continues as $P$. For a guarded choice $P_1 \,\Vert_{g}\, P_2$ with $g \subseteq S$, the process behaves as $P_1$ when the current state $s \in g$, and as $P_2$ otherwise.

\subsection{Vertex Conflict Avoidance Example with the Pipeline}

\begin{figure}[H]
  \centering
  \begin{tikzpicture}[scale=0.7]
    \draw[step=1cm, gray!40, very thin] (0,0) grid (3,3);

    \fill[gray!35] (0,2) rectangle (1,3);
    \fill[gray!35] (2,2) rectangle (3,3);
    \fill[gray!35] (0,0) rectangle (1,1);
    \fill[gray!35] (2,0) rectangle (3,1);

    \fill[blue!40] (1,2) rectangle (2,3);
    \node at (1.5,2.5) {$A_1$};
    \fill[red!35] (0,1) rectangle (1,2);
    \node at (0.5,1.5) {$A_2$};

    \draw[blue, thick] (1,0) rectangle (2,1);
    \node[blue] at (1.5,0.5) {$T_1$};
    \draw[red, thick] (2,1) rectangle (3,2);
    \node[red] at (2.5,1.5) {$T_2$};

  \end{tikzpicture}
  \caption{Initial state for the ``Blind Agents'' ($R=0$).}
  \label{fig:mot-env}
\end{figure}
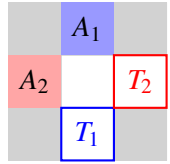

\noindent An interesting motivating example arises when the agents are blind ($R = 0$), a scenario we refer to as the ``Blind Agents'', in the environment shown in Figure~\ref{fig:mot-env}.

\noindent Since each agent's local observation remains constant over time, it is impossible to construct a shield that depends solely on local observations \cite{melcer2024shielddecentralization}. Likewise, DFA-based constructions that attempt to avoid unsafe states \cite{elsayedaly2021safemultiagentreinforcementlearning} are ineffective in this setting, because no action available to any agent can be guaranteed to lead, in combination, to a safe joint action.

By contrast, our method enriches the available information through the process syntax, hence it is able to address this problem using the pipeline as shown below:

\paragraph{Pipeline Initiation} To instantiate the pipeline, we provide a transition description $SAS:S\times S\to 2^A$, an observation description $O'_i:S\to 2^{\Omega_i}$, and an initial set description $S_0\subseteq S$. These descriptions may be instantiated as the supports of $\delta$, $O$ and $\rho_0$, or as conservative abstractions thereof.

\paragraph{Shield Process Description} We begin with the following process specification:

\[P = (Sh_1.Sh_2.Sh_3.\textsf{idle})\Vert_g\textsf{fail}\]

where $g = \{\begin{tikzpicture}[scale=0.5, baseline=(current bounding box.center)]
    \draw[step=1cm, gray!40, very thin] (0,0) grid (3,3);

    \fill[gray!35] (0,2) rectangle (1,3);
    \fill[gray!35] (2,2) rectangle (3,3);
    \fill[gray!35] (0,0) rectangle (1,1);
    \fill[gray!35] (2,0) rectangle (3,1);

    \fill[blue!40] (1,2) rectangle (2,3);
    \node at (1.5,2.5) {$A_1$};
    \fill[red!35] (0,1) rectangle (1,2);
    \node at (0.5,1.5) {$A_2$};

    \draw[blue, thick] (1,0) rectangle (2,1);
    \node[blue] at (1.5,0.5) {$T_1$};
    \draw[red, thick] (2,1) rectangle (3,2);
    \node[red] at (2.5,1.5) {$T_2$};
    \node[red] at (2.5,1.5) {$T_2$};

  \end{tikzpicture}\}$, $Sh_1 = \{\begin{tikzpicture}[scale=0.5, baseline=(current bounding box.center)]
    \draw[step=1cm, gray!40, very thin] (0,0) grid (3,3);

    \fill[gray!35] (0,2) rectangle (1,3);
    \fill[gray!35] (2,2) rectangle (3,3);
    \fill[gray!35] (0,0) rectangle (1,1);
    \fill[gray!35] (2,0) rectangle (3,1);

    \fill[blue!40] (1,1) rectangle (2,2);
    \node at (1.5,1.5) {$A_1$};
    \fill[red!35] (0,1) rectangle (1,2);
    \node at (0.5,1.5) {$A_2$};

    \draw[blue, thick] (1,0) rectangle (2,1);
    \node[blue] at (1.5,0.5) {$T_1$};
    \draw[blue, thick] (1,0) rectangle (2,1);
    \draw[red, thick] (2,1) rectangle (3,2);
    \node[red] at (2.5,1.5) {$T_2$};

  \end{tikzpicture}\}$, $Sh_2 = \{\begin{tikzpicture}[scale=0.5, baseline=(current bounding box.center)]
    \draw[step=1cm, gray!40, very thin] (0,0) grid (3,3);

    \fill[gray!35] (0,2) rectangle (1,3);
    \fill[gray!35] (2,2) rectangle (3,3);
    \fill[gray!35] (0,0) rectangle (1,1);
    \fill[gray!35] (2,0) rectangle (3,1);

    \fill[blue!40] (1,0) rectangle (2,1);
    \node at (1.5,0.5) {$A_1$};
    \fill[red!35] (1,1) rectangle (2,2);
    \node at (1.5,1.5) {$A_2$};

    \draw[blue, thick] (1,0) rectangle (2,1);
    \draw[red, thick] (2,1) rectangle (3,2);
    \node[red] at (2.5,1.5) {$T_2$};

  \end{tikzpicture}\}$, $Sh_3 = \{\begin{tikzpicture}[scale=0.5, baseline=(current bounding box.center)]
    \draw[step=1cm, gray!40, very thin] (0,0) grid (3,3);

    \fill[gray!35] (0,2) rectangle (1,3);
    \fill[gray!35] (2,2) rectangle (3,3);
    \fill[gray!35] (0,0) rectangle (1,1);
    \fill[gray!35] (2,0) rectangle (3,1);

    \fill[blue!40] (1,0) rectangle (2,1);
    \node at (1.5,0.5) {$A_1$};
    \fill[red!35] (2,1) rectangle (3,2);
    \node at (2.5,1.5) {$A_2$};

    \draw[blue, thick] (1,0) rectangle (2,1);
    \draw[red, thick] (2,1) rectangle (3,2);

  \end{tikzpicture}\}$
  
\noindent demonstrates a sequence of safe states that ensures both agents eventually reach their respective targets without encountering vertex conflicts.

\begin{figure}[ht]
\centering
\begin{tikzpicture}[
  >=Stealth, semithick,
  st/.style={draw, rounded corners, inner sep=2pt, align=center, font=\scriptsize},
  lab/.style={font=\scriptsize},
  node distance=2.3cm
]
  \node[st] (start) {$\textsf{start}$};
  \node[st, right=of start] (p1) {$Sh_1.Sh_2.Sh_3.\textsf{idle}$};
  \node[st, right=of p1] (p2) {$Sh_2.Sh_3.\textsf{idle}$};
  \node[st, right=of p2] (p3) {$Sh_3.\textsf{idle}$};
  \node[st, right=of p3] (idle) {$\textsf{idle}$};
  \node[st, below=1.0cm of p1] (fail) {$\textsf{fail}$};

  \draw[->] (start) -- node[lab,above] {$g$} (p1);
  \draw[->] (start) -- node[lab,left] {$S\backslash g$} (fail);
  \draw[->] (p1) -- node[lab,left] {$S\backslash Sh_1$} (fail);
  \draw[->] (p1) -- node[lab,above] {$Sh_1$} (p2);
  \draw[->] (p2) -- node[lab,left] {$S\backslash Sh_2$} (fail);
  \draw[->] (p2) -- node[lab,above] {$Sh_2$} (p3);
  \draw[->] (p3) -- node[lab,left] {$S\backslash Sh_3$} (fail);
  \draw[->] (p3) -- node[lab,above] {$Sh_3$} (idle);
  \draw[->] (fail) edge[loop below] node[lab] {$S$} (fail);
  \draw[->] (idle) edge[loop above] node[lab] {$S$} (idle);
\end{tikzpicture}
\caption{Shield process automaton for the ``Blind Agents''.}
\label{fig:mot-automata}
\end{figure}
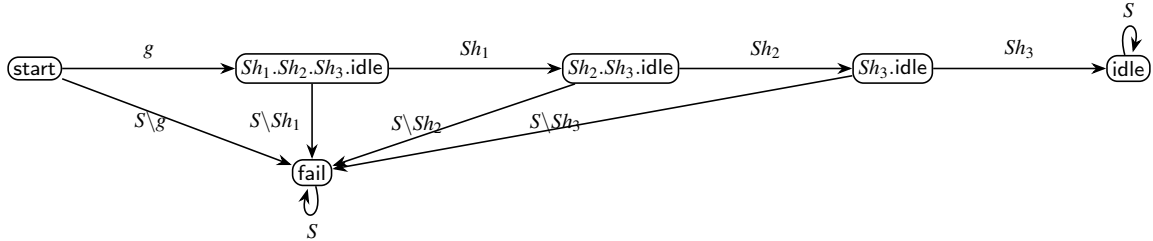

\paragraph{Shield Process Automaton} We generate a shield process automaton from the process specification, obtaining a deterministic automaton that interprets the shield process:
\[
  \mathcal{A}_{P_0}=(Q_P,\Sigma,\delta_P,q_0),
  \qquad
  \Sigma:=S,\quad q_0:=\textsf{start}
\]
\noindent Its states are either $\textsf{idle}$, $\textsf{fail}$, a continuation process (e.g., $Sh_2.Sh_3.\textsf{idle}$), or the dummy initial state $\textsf{start}$. Transitions follow the guard set $g$ and consume the global state shield $Sh$ in sequence. The resulting automaton for the ``Blind Agents'' is shown in Figure~\ref{fig:mot-automata}.

\paragraph{Global Shield Mealy Machine} We construct the global shield Mealy machine
\[
  G=(Q_G,q^G_0,\Sigma^G_I,\Sigma^G_O,\delta_G,\lambda_G),
\]
where
$Q_G := \{\textsf{idle},\textsf{fail}\}\ \cup\ (2^{S}\times (Q_P\backslash\{\textsf{idle},\textsf{fail}\}))$,
$q^G_0 := (S_0,\textsf{start})$,
$\Sigma^G_I := S$, and
$\Sigma^G_O := \bigl(\prod_{i=1}^n 2^{A_i}\bigr)\ \cup\ \{\bot\}$.
Here, the distinguished output symbol $\bot$ denotes shield failure: no safe action set can be provided for the current situation.
Each global Mealy state thus records the current set of reachable Dec-POMDP states together with the current state of the process automaton.

To compute outputs, we fix a deterministic decomposition
$\textsf{Dec}:2^A \to \prod_{i=1}^n 2^{A_i}$ that turns a joint action set into one safe local action set per agent. For
$\textsf{Dec}(\hat{A})=(U_1,\dots,U_n)$, where $\hat{A}\subseteq A$ is a set of joint actions, we require
$\prod_{i=1}^n U_i \subseteq \hat{A}$ and
$\prod_{i=1}^n U_i=\emptyset \Longleftrightarrow \hat{A}=\emptyset$.
Thus, if each agent independently chooses an action from its local set, the resulting joint action remains in $\hat{A}$.
In this paper, we instantiate $\textsf{Dec}$ by selecting, in a deterministic manner, a maximum-cardinality product set $\prod_{i=1}^n U_i \subseteq \hat{A}$.

For a transition from $(S_{\mathsf{cur}},q)$ to $(S_{\mathsf{next}},q')$, we compute
$\hat{A}$ by intersecting, over all $s\in S_{\mathsf{cur}}\cap g$, with $q\xrightarrow{g}q'$, the joint actions that can move
from $s$ into any state in $S_{\mathsf{next}}$ using $SAS$, and output $\textsf{Dec}(\hat{A})$.

\begin{figure}[ht]
\centering
\begin{tikzpicture}[
  >=Stealth, semithick,
  st/.style={draw, rounded corners, inner sep=2pt, align=center, font=\scriptsize},
  lab/.style={font=\scriptsize},
  node distance=2.4cm
]
  \node[st] (q0) {$(g,\textsf{start})$};
  \node[st, right=of q0] (q1) {$(Sh_1,Sh_1.Sh_2.Sh_3.\textsf{idle})$};
  \node[st, right=of q1] (q2) {$(Sh_2,Sh_2.Sh_3.\textsf{idle})$};
  \node[st, right=of q2] (q3) {$(Sh_3,Sh_3.\textsf{idle})$};
  \node[st, below=1.0cm of q3] (idle) {$\textsf{idle}$};
  \node[st, below=1.0cm of q1] (fail) {$\textsf{fail}$};

  \draw[->] (q0) -- node[lab,above] {$g\ /\ [\downarrow_{A_1},\cdot_{A_2}]$} (q1);
  \draw[->] (q0) -- node[lab,left] {$S\backslash g\ /\ \bot$} (fail);
  \draw[->] (q1) -- node[lab,left] {$S\backslash Sh_1\ /\ \bot$} (fail);
  \draw[->] (q1) -- node[lab,above] {$Sh_1\ /\ [\downarrow_{A_1},\rightarrow_{A_2}]$} (q2);
  \draw[->] (q2) -- node[lab,left] {$S\backslash Sh_2\ /\ \bot$} (fail);
  \draw[->] (q2) -- node[lab,above] {$Sh_2\ /\ [\cdot_{A_1},\rightarrow_{A_2}]$} (q3);
  \draw[->] (q3) -- node[lab,below] {$S\backslash Sh_3\ /\ \bot$} (fail);
  \draw[->] (q3) -- node[lab,right] {$Sh_3\ /\ [\cdot_{A_1},\cdot_{A_2}]$} (idle);
  \draw[->] (fail) edge[loop below] node[lab] {$\bot$} (fail);
  \draw[->] (idle) edge[loop below] node[lab] {$S$} (idle);
\end{tikzpicture}
\caption{Global shield Mealy machine for the ``Blind Agents''.}
\label{fig:mot-global-mealy}
\end{figure}
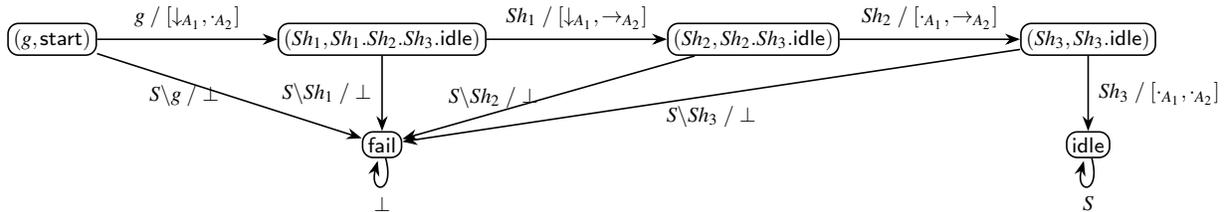

For the ``Blind Agents'', the resulting global shield Mealy machine is shown in Figure~\ref{fig:mot-global-mealy}.

\paragraph{Local Shield Mealy Machine}
The local shield for agent $i$ is a Mealy machine
\[
  L_i=(Q_{L_i},q^{L_i}_0,\Sigma^{L_i}_I,\Sigma^{L_i}_O,\delta_{L_i},\lambda_{L_i}),
\]
where
$Q_{L_i}:=2^{Q_G}$,
$q^{L_i}_0:=\{(S_0,\textsf{start})\}$,
$\Sigma^{L_i}_I:=\Omega_i$, and
$\Sigma^{L_i}_O:=2^{A_i}\cup\{\bot\}$.
Each local Mealy state is a belief subset over global Mealy states consistent with agent $i$'s current observation, as determined by the observation description $O'_i$. The transition function is defined by $
  \delta_{L_i}(q^{L_i},o_i)
  :=
    \bigcup_{(\hat{S}, q)\in q^{L_i}}\ \{\, \delta_G((\hat{S}, q),s)\mid s\in \hat{S},\ o_i\in O'_i(s) \,\}
$, where $q^{L_i}\subseteq Q_G$ is the current local Mealy state and $o_i\in \Omega_i$ is the observation. 

The local Mealy output is obtained by intersecting the $i$th components of the global Mealy outputs over the global Mealy states represented by the current local Mealy state. We treat $\textsf{fail}$ as contributing all actions unless every state in the belief is $\textsf{fail}$, in which case the output is $\bot$, indicating a shield failure: the shield cannot provide any action set that is guaranteed to be safe for the agent to choose from.

In the ``Blind Agents'', all observations are identical, so the belief state remains maximal and all process branches are treated as possible. The resulting local shields are shown in Figure~\ref{fig:mot-local-mealy}.

\newcommand{\motQzeroA}{\ensuremath{\{(g,\textsf{start})\}}}
\newcommand{\motQzeroB}{\ensuremath{\{(g,\textsf{start})\}}}
\newcommand{\motQone}{\ensuremath{\{(Sh_1,Sh_1.Sh_2.Sh_3.\textsf{idle}),\textsf{fail}\}}}
\newcommand{\motQtwo}{\ensuremath{\{(Sh_2,Sh_2.Sh_3.\textsf{idle}),\textsf{fail}\}}}
\newcommand{\motQthree}{\ensuremath{\{(Sh_3,Sh_3.\textsf{idle}),\textsf{fail}\}}}
\newcommand{\motQfour}{\ensuremath{\{\textsf{idle},\textsf{fail}\}}}

\begin{figure}[H]
  \centering
  \begin{subfigure}[t]{0.48\linewidth}
    \centering
    \begin{tikzpicture}[
      >=Stealth, thick,
      st/.style={draw, rounded corners, inner sep=1.5pt, align=center, font=\scriptsize},
      lab/.style={font=\scriptsize}
    ]
      \def\dx{3.8}
      \def\dy{2.0}

      \node[st] (a1) at (0,0) {\motQzeroA};
      \node[st] (a2) at (\dx,0) {\motQone};
      \node[st] (a3) at (\dx,-\dy) {\motQtwo};
      \node[st] (a4) at (0,-\dy) {\motQthree};
      \node[st] (a5) at (0,-2*\dy) {\motQfour};

      \draw[->] (a1) -- node[lab,above] {$\cdot/\{\downarrow\}$} (a2);
      \draw[->] (a2) -- node[lab,right] {$\cdot/\{\downarrow\}$} (a3);
      \draw[->] (a3) -- node[lab,below] {$\cdot/\{\cdot\}$} (a4);
      \draw[->] (a4) -- node[lab,right] {$\cdot/\{\cdot\}$} (a5);
      \draw[->] (a5) edge[loop below] node[lab] {$\cdot/\{\cdot\}$} (a5);
    \end{tikzpicture}
    \caption{Local shield for $A_1$.}
  \end{subfigure}
  \hfill
  \begin{subfigure}[t]{0.48\linewidth}
    \centering
    \begin{tikzpicture}[
      >=Stealth, thick,
      st/.style={draw, rounded corners, inner sep=1.5pt, align=center, font=\scriptsize},
      lab/.style={font=\scriptsize}
    ]
      \def\dx{3.8}
      \def\dy{2.0}

      \node[st] (b1) at (0,0) {\motQzeroB};
      \node[st] (b2) at (\dx,0) {\motQone};
      \node[st] (b3) at (\dx,-\dy) {\motQtwo};
      \node[st] (b4) at (0,-\dy) {\motQthree};
      \node[st] (b5) at (0,-2*\dy) {\motQfour};

      \draw[->] (b1) -- node[lab,above] {$\cdot/\{\cdot\}$} (b2);
      \draw[->] (b2) -- node[lab,right] {$\cdot/\{\rightarrow\}$} (b3);
      \draw[->] (b3) -- node[lab,below] {$\cdot/\{\rightarrow\}$} (b4);
      \draw[->] (b4) -- node[lab,right] {$\cdot/\{\cdot\}$} (b5);
      \draw[->] (b5) edge[loop below] node[lab] {$\cdot/\{\cdot\}$} (b5);
    \end{tikzpicture}
    \caption{Local shield for $A_2$.}
  \end{subfigure}
\caption{Local shield Mealy machines in the "Blind Agents".}
\label{fig:mot-local-mealy}
\end{figure}
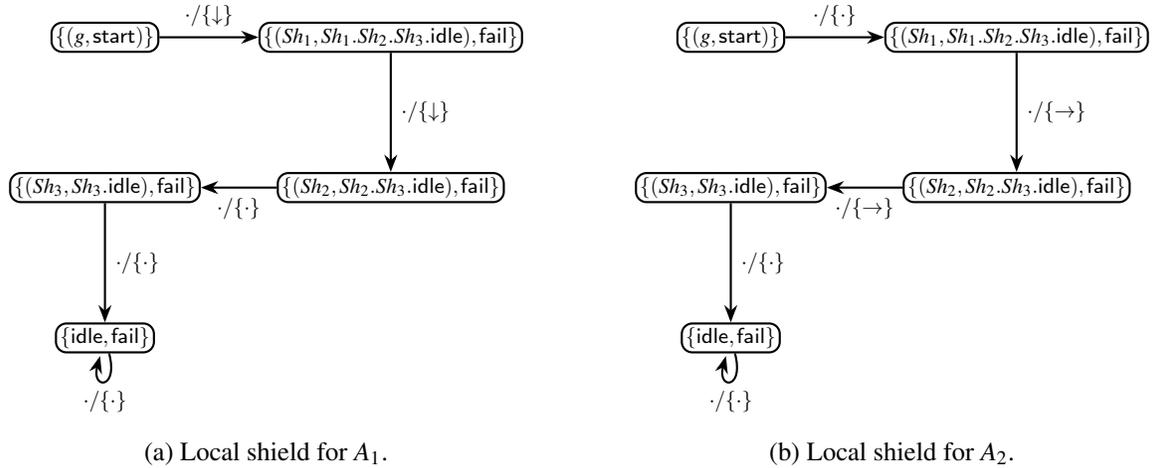

\paragraph{PRISM analysis} We analyse the shielded model using PRISM model checker, treating the agents' policies as nondeterministic. PRISM computes lower and upper bounds on both the probability of shield failure and the probability of reaching unsafe states; in this instance, $\textsf{shield failure}$ has lower and upper bounds of $0.000000$, and $\textsf{reaching unsafe states}$ has lower and upper bounds of $0.000000$, indicating that our shield ends correctly with no shield failure or reaching unsafe states.

\section{Case Study}

In this case study, we compare multiple shield process specifications in a more general MAPF setting, using the unshielded case as the baseline.

We evaluate random instances with $n = 2$ and $n = 3$ agents on $3 \times 3$, $4 \times 4$, and $5 \times 5$ grids, using the following shield process specifications:
\[
\begin{aligned}
P_1 &= \mu X.S_{safe}.X\\
P_2 &= \mu X.(S_{safe}.X\Vert_{g_{o_1}}(S_{safe}.X\Vert_{g_{o_2}}(\cdots(S_{safe}.X\Vert_{g_{o_{\cdots}}}\textsf{fail}))))
\end{aligned}
\]
where $S_{\mathit{safe}}$ ranges over safe states, and $g_o$ enumerates all global states consistent with each joint observation $o\in\Omega$.

Process $P_1$ is a conservative shield that primarily aims to keep the system in safe states; to this end, it often forces each agent to remain in its original position.

Process $P_2$ is the least conservative shield obtainable under our current pipeline. It minimises the loss of actions due to intersections introduced by the branch structure (i.e., it avoids disallowing multiple partial observations whenever possible) and, conversely, admits the largest possible set of next safe states.

In addition to the number of agents and the grid size, our case study also varies several environmental and evaluation parameters. Specifically, for each configuration we keep the number of obstacles fixed, since obstacle density is associated with the probability of vertex conflicts. We then evaluate each run along three dimensions: $\textsf{collision}$ (vertex conflicts), $\textsf{shield failure}$ ($\bot$ outputs from the local Mealy machines), and $\textsf{reached}$ (all agents reaching their targets).

\begin{table}[ht]
  \centering
  \small
  \begin{tabular}{cccc l c c c}
    \hline
    grid & $n$ & $R$ & obstacles & shield & $\textsf{collision}$ & \textsf{shield failure} & $\textsf{reached}$ \\ \hline
    $4\times 4$ & 2 & $-$ & 9 & $/$ & 0.727 & 0.000 & 0.273\\
    $4\times 4$ & 2 & 2 & 9 & $P_1$ & 0.000 & 0.000 & 0.051\\
    $4\times 4$ & 2 & 2 & 9 & $P_2$ & 0.000 & 0.000 & 0.929\\
    \hline
    $4\times 4$ & 2 & 1 & 9 & $P_1$ & 0.000 & 0.000 & 0.055\\
    $4\times 4$ & 2 & 1 & 9 & $P_2$ & 0.000 & 0.000 & 0.906\\
    \hline
    $4\times 4$ & 2 & 0 & 9 & $P_1$ & 0.000 & 0.000 & 0.046\\
    $4\times 4$ & 2 & 0 & 9 & $P_2$ & 0.000 & 0.325 & 0.295\\
    \hline
    $3\times 3$ & 3 & $-$ & 3 & $/$ & 0.986 & 0.000 & 0.015\\
    $3\times 3$ & 3 & 1 & 3 & $P_1$ & 0.000 & 0.000 & 0.017\\
    $3\times 3$ & 3 & 1 & 3 & $P_2$ & 0.000 & 0.363 & 0.292\\
    \hline
    $4\times 4$ & 3 & $-$ & 9 & $/$ & 0.969 & 0.000 & 0.031\\
    $4\times 4$ & 3 & 1 & 9 & $P_1$ & 0.000 & 0.000 & 0.006\\
    $4\times 4$ & 3 & 1 & 9 & $P_2$ & 0.000 & 0.414 & 0.256\\
    \hline
    $4\times 4$ & 3 & 2 & 9 & $P_1$ & 0.000 & 0.000 & 0.009\\
    $4\times 4$ & 3 & 2 & 9 & $P_2$ & 0.000 & 0.022 & 0.533\\
    \hline
  \end{tabular}
  \caption{MAPF experiments under a random policy.}
  \label{tab:benchmark}
\end{table}
We first benchmark the different shield process specifications under a random policy across some MAPF configurations (Table~\ref{tab:benchmark}). Here, a random policy means that each agent samples uniformly from its currently available actions; in the shielded cases, this uniform sampling is taken over the action set returned by the shield.

In Table~\ref{tab:benchmark}, across all configurations, the unshielded random policy yields high $\textsf{collision}$, as it does not discriminate between colliding and non-colliding actions. In contrast, both shields eliminate vertex conflicts in all configurations, providing a strict safety improvement over the unshielded baseline.

Among the shielded, $P_2$ consistently attains a higher $\textsf{reached}$ than both $P_1$ and the unshielded model. $P_2$ removes colliding joint actions while still admitting many safe moves, so agents can explore more states and execute more actions under a random policy. By comparison, the conservatism of $P_1$ over-suppresses many potential safe actions.

Varying the observation radius $R$ shows that a larger observation radius improves the performance of $P_2$. As $R$ increases, $P_2$ achieves higher reachability, since richer observations enable more precise belief updates and thus larger safe action sets. \textsf{shield failure} occurs in configurations where (e.g., $n = 3$, $R = 1$ with dense obstacles), limited observability prevents the shield from always proposing a non-empty safe joint action. As $R$ increases, \textsf{shield failure} decreases rapidly.

\begin{figure}[H]
  \centering
  \includegraphics[width=\linewidth]{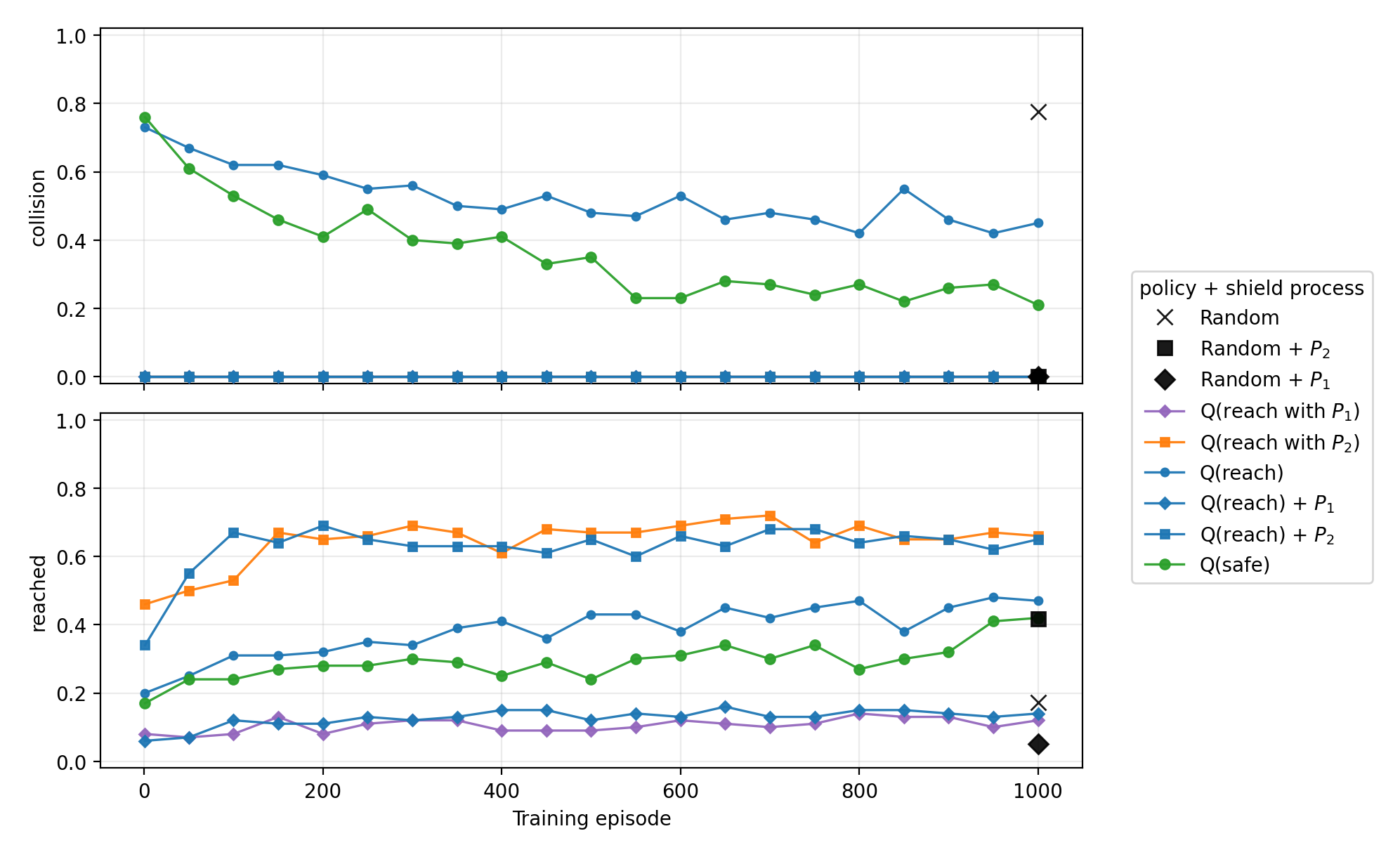}
  \caption{Evaluation curves under training for learned policies with shields under $3\times 3$, $n=2$, $R=1$ configuration}
  \label{fig:train}
\end{figure}

We next evaluate four policies trained using Q-learning with Q-tables \cite{10.1007/BF00992698} in a $3 \times 3$ MAPF configuration with uniformly random obstacle numbers across training and test environments (Figure~\ref{fig:train}). Q-learning is a reinforcement-learning algorithm that updates action values from sampled transitions, and a Q-table is the finite table storing those state-action values. The policies are: $Q_{\text{reach}}$, trained to maximise the probability of reaching targets only; $Q_{\text{safe}}$, trained to maximise target reach while explicitly penalising vertex conflicts in the reward function; $Q_{\text{reach with }P_1}$, trained to maximise reach, with shield process $P_1$ enforced during interaction; and $Q_{\text{reach with }P_2}$, trained to maximise reach, with shield process $P_2$ enforced.

For comparison, the figure also includes random policies (unshielded, shielded by $P_1$, and shielded by $P_2$) as baselines, as well as the post-hoc shielding of $Q_{\text{reach}}$ by $P_1$ and by $P_2$. $\textsf{shield failure}$ does not occur in this configuration and is therefore omitted from the plots. Black dots indicate the random baselines.

The learning curves show that both $Q_{\text{reach}}$ and $Q_{\text{safe}}$ exhibit non-zero $\textsf{collision}$, but these rates decrease over training. $\textsf{collision}$ declines faster for $Q_{\text{safe}}$, as expected, due to the explicit vertex conflicts penalty. All shielded variants maintain zero $\textsf{collision}$ throughout, demonstrating that the shields enforce safety regardless of the underlying policy. In terms of $\textsf{reached}$, $Q_{\text{safe}}$ attains slightly lower success rates than $Q_{\text{reach}}$, reflecting the trade-off introduced by the vertex conflicts penalty in learning.

The best performance in $\textsf{reached}$ is obtained by $Q_{\text{reach with }P_2}$ and by shielding $Q_{\text{reach}}$ with $P_2$. These combinations achieve high $\textsf{reached}$ while maintaining zero $\textsf{collision}$, showing that $P_2$ provides strong safety guarantees without unduly restricting actions. In contrast, policies trained with $P_1$ or evaluated under $P_1$ perform only slightly better than the random policy with $P_1$: the over-conservatism of $P_1$ severely limits movement, so even a trained policy cannot substantially exceed the baseline, though it may still discover narrow paths to some targets. Overall, these results reinforce the earlier findings: $P_2$ delivers a more favourable safety-performance trade-off than $P_1$, and all shielding can strictly improve safety relative to the unshielded while preserving, or even enhancing, task performance across environments.

\section{Related Work}

Bloem et al.\ \cite{bloem2015shield_synthesis_runtime_enforcement} establish reactive-synthesis and temporal-logic foundations that underpin many runtime-enforcement and shield-style artefacts. Alshiekh et al.\ \cite{alshiekh2017safereinforcementlearningshielding} propose temporal-logic shields for safe reinforcement learning by synthesising an online action filter that blocks unsafe actions during learning and execution. Jansen et al.\ \cite{jansen2020safe_rl_probabilistic_shields} propose probabilistic shields that use probabilistic model checking to restrict actions so that safety objectives hold with explicit probability guarantees. Elsayed-Aly et al.\ \cite{elsayedaly2021safemultiagentreinforcementlearning} extend shielding to multi-agent reinforcement learning by enforcing safety properties in settings where multiple agents jointly determine outcomes.
Melcer et al.\ \cite{melcer2022shield_decentralization_neurips} introduce shield decentralisation by decomposing a centralised safe-action filter into per-agent filters suitable for decentralised execution. Melcer et al.\ \cite{melcer2024shielddecentralization} study decentralisation under general partial observability via a SAT-based formulation that decides whether local action sets can be chosen so that their product remains within the centralised safe joint-action set. Melcer et al.\ \cite{Melcer2025LearnedShields} explore learned shields for multi-agent reinforcement learning by training shield policies when explicit models or specifications are unavailable or costly to engineer.
El~Mqirmi et al.\ \cite{elmqirmi2021abstraction_based_checking_marl} propose abstraction-based post-hoc checking for deep multi-agent reinforcement learning, verifying learned behaviours against desired properties rather than filtering actions online.

\section{Conclusion and Future Work}
This paper presents a shield process algebra for specifying safe global behaviour under Dec-POMDP, together with a compilation pipeline that produces local Mealy shields executable from local observations. The resulting local Mealy shields enforce safety without communication or shared global states and are implemented in Rust with PRISM integration for model analysis. Our case study illustrates how the algebra and pipeline can eliminate collisions while preserving task performance.

Future work includes optimising the decentralised joint-action decomposition (\textsf{Dec}), for example by integrating SAT-based decentralisation methods \cite{melcer2024shielddecentralization}; combining our approach with learned shields \cite{Melcer2025LearnedShields} to reduce manual specification effort; and improving robustness through probabilistic shield design \cite{jansen2020safe_rl_probabilistic_shields}. Another direction is to extend the algebra with compositional constructs, such as combinations of shield processes, to improve both semantic clarity and pipeline efficiency.
\section*{Acknowledgements}
\begin{sloppypar}
The first author received travel support to attend PLACES 2026 from Balliol College, University of Oxford, through the Donald Michie Scholarship Fund.

The second author is partially supported by EPSRC grants EP/T006544/2, EP/T014709/2, EP/Y005244/1, EP/V000462/1, EP/X015955/1, EP/Z0005801/1; Horizon EU TaRDIS 101093006 (UKRI No.~10066667); and ARIA.
\end{sloppypar}

\bibliographystyle{eptcs}
\bibliography{references}

\end{document}